\documentclass[aps,prl,twocolumn,superscriptaddress,showpacs]{revtex4}

\newcommand{\apf}{$\overline{\mbox{Pf}}\;$}

\usepackage{graphicx} 

\begin{document}
\title{Spontaneous Particle-Hole Symmetry Breaking
in the $\nu=5/2$ Fractional Quantum Hall Effect}

\author{Michael R. Peterson}
\affiliation{Condensed Matter Theory Center, Department of
Physics, University of Maryland, College Park, MD 20742}

\author{Kwon Park}
\affiliation{School of Physics, Korea Institute for Advanced Study,
Seoul 130-722, Korea}

\author{S.  Das  Sarma}
\affiliation{Condensed Matter Theory Center, Department of
Physics, University of Maryland, College Park, MD 20742}

\date{\today}

\begin{abstract}
The essence of the $\nu=5/2$ fractional quantum Hall effect is
believed to be captured by the Moore-Read Pfaffian (or anti-Pfaffian)
description.  However, a mystery regarding the formation of the
Pfaffian state is the role of the three-body interaction Hamiltonian
$H_3$ that produces it as an exact ground state and the concomitant
particle-hole symmetry breaking.  We show that a two-body interaction
Hamiltonian $H_2$ constructed via particle-hole symmetrization of
$H_3$ produces a ground state nearly exactly approximating the
Pfaffian and anti-Pfaffian states, respectively, in the spherical
geometry.  Importantly, the ground state energy of $H_2$ exhibits a
``Mexican-hat'' structure as a function of particle number in the
vicinity of half filling for a given flux indicating spontaneous
particle-hole symmetry breaking.  This signature is absent for the
second Landau level Coulomb interaction at 5/2.
\end{abstract}


\maketitle

The fractional quantum Hall effect (FQHE)~\cite{fqhe} at orbital Landau
level (LL) filling factor $\nu=5/2$~\cite{willett} (1/2
filling of the second LL (SLL)) is the subject of recent
theoretical and experimental research. This is partly due to
the Moore-Read Pfaffian state~\cite{mr}, the leading
theoretical description of the $\nu=5/2$ FQHE, possessing
non-Abelian quasiparticle excitations with potential applications 
towards fault-tolerant 
topological quantum computation~\cite{tqc}.  Recent theoretical
results~\cite{ft-long,ft-short} along with previous
work~\cite{morf,hr} provide compelling evidence that this
non-abelian description is essentially correct.

However, a question remains regarding the
Moore-Read Pfaffian (Pf) description best illustrated by contrasting
it to the celebrated Laughlin state~\cite{laugh} for the FQHE
at $\nu=1/q$ ($q$ odd). When confined to a single LL,
two-body interaction Hamiltonians can be 
parameterized by Haldane pseudopotentials $V_m$--the
energy for a pair of electrons in a state of relative angular
momentum $m$~\cite{sphere} where only odd $m$ enters for
spin-polarized electron systems~\cite{foot}. The Laughlin
state is the exact ground state of a two-body Hamiltonian 
with only $V_1$ non-zero (the
interaction is hard-core).  Thus, through the pseudopotential
description, the Laughlin state is shown to be continuously connected to the
exact ground state of the Coulomb Hamiltonian at $\nu=1/q$.  The
Pf wave function, by contrast, is an exact ground state of a repulsive
three-body Hamiltonian $H_3$ for even number of electrons $N_e$ 
in a half-filled LL~\cite{gww}.  There is no 
two-body Hamiltonian, and hence no exact pseudopotential description, 
for which the Pf is an exact eigenstate.  So, as
good as the physical description for the $\nu=5/2$ FQHE state
provided by the Pf may be, it is not continuously connected to the
exact Coulomb ground state or, in fact, the ground state of {\it any} two-body
Hamiltonian.  This notion has been discussed in the
literature~\cite{ho,nayak,read} for over ten years, and recently 
questions have been raised~\cite{jain,jain1} about the applicability of the Pf 
for the physical 5/2-state.

However, evidently some two-body Hamiltonians produce
ground states that have nearly unity overlap ($\approx0.99$) with
the Pf.  For example, Morf~\cite{morf} showed in the spherical
geometry (see below for details) that, for $N_e=8$, the SLL Coulomb
Hamiltonian has a nearly Pf ground state if $V_1\rightarrow
1.1V_1$, i.e., $V_1$ is slightly increased from its SLL Coulomb value. In the torus
geometry, Rezayi-Haldane~\cite{hr} showed a nearly Pf ground state
(although they compared to a particle-hole symmetrized Pf state)
for an increase in $V_1$ and/or a decrease in $V_3$.  A
conceptual problem with these results is that no physical
effect can produce an increased $V_1$--although a decreased
$V_3$ is possible. Recently, it was
shown~\cite{ft-short,ft-long} (sphere and torus)
that the inclusion of finite-thickness effects inherent in
realistic experimental quantum wells produces a nearly Pf ground
state for well widths of about four magnetic lengths. Clearly, however,
in realistic calculations {\em all} $V_m$'s change (not 
only $V_1$, $V_3$).

To make matters more interesting is the recently discussed
fact~\cite{apf1,apf2} that the Pf is not particle-hole (PH)
symmetric since it is the exact ground state of a
three-body interaction Hamiltonian that explicitly breaks PH
symmetry. Consequently, if correct, the Pf description of the
$\nu=5/2$ FQHE would require spontaneous PH symmetry breaking of the 
actual two-body Coulombic Hamiltonian--in
the absence of inter-LL mixing which could break PH symmetry in real systems 
explicitly. This observation, in turn, leads to an
identification of the PH conjugate state, the anti-Pfaffian
(\apf), that is degenerate with the Pf in the PH symmetry
respecting limit.  An important question is whether PH
symmetry is indeed broken spontaneously without LL
mixing and addressing this question is a main goal of this
work.

We begin our quantitative analysis by constructing a two-body
interaction Hamiltonian $H_2$ which is PH symmetric and yet
contains as much of the physics of $H_3$ as possible. The
purpose is to use $H_2$ as a reference model Hamiltonian to which
other more realistic Hamiltonians are compared. First
note that, in the spherical geometry, $H_3=
\sum_{i<j<k} {\cal P}_{ijk}(3N_\phi/2-3)$ with ${\cal P}_{ijk}(L)$
projecting onto electron triplets with total angular momentum $L$ and
$N_\phi$ the total magnetic flux piercing the
surface.  We consider a PH transformation of $H_3$ by
taking $c_m^\dagger$ $(c_m)$ $\rightarrow$ $c_m$ $(c_m^\dagger)$
where $c_m^\dagger$ creates an electron in an angular
momentum state $m$. We call this PH-conjugate Hamiltonian
$\overline{H}_3$ and, naturally, the {\apf} is its 
ground state. Importantly, when normal
ordered, $\overline{H}_3$ contains a three-body term 
exactly the minus of $H_3$ plus a two-body term, a
one-body term (the chemical potential), and a
constant~\cite{apf1}.  Adding $H_3$
and $\overline{H}_3$ simply eliminates the three-body interaction 
and simultaneously restores the PH symmetry suggesting an
intriguing possibility: $H_2 \equiv H_3 + \overline{H}_3$.

\begin{figure}[t]
\begin{center}
\mbox{\includegraphics[width=5.1cm,angle=-90]{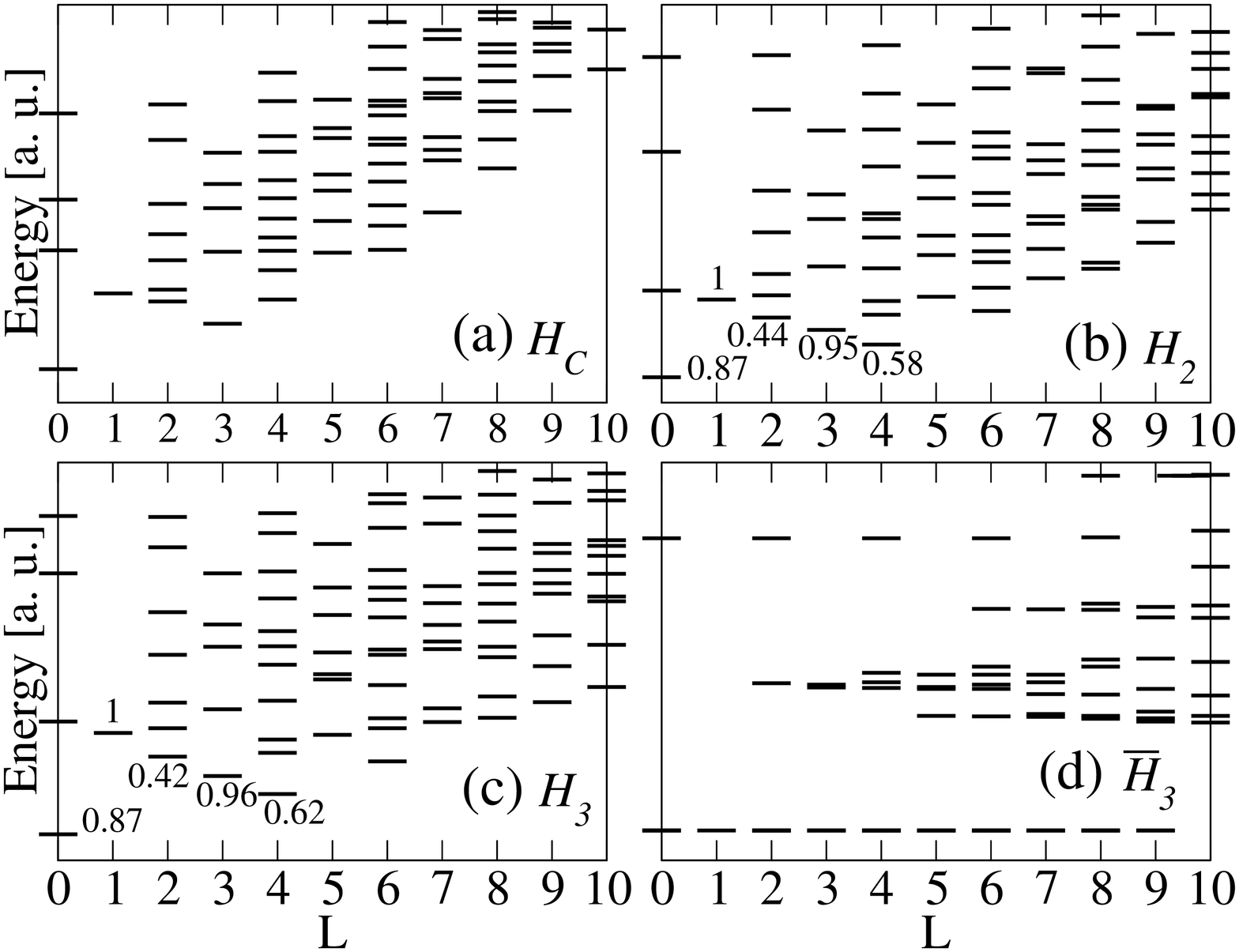}}\\
\mbox{\includegraphics[width=5.1cm,angle=-90]{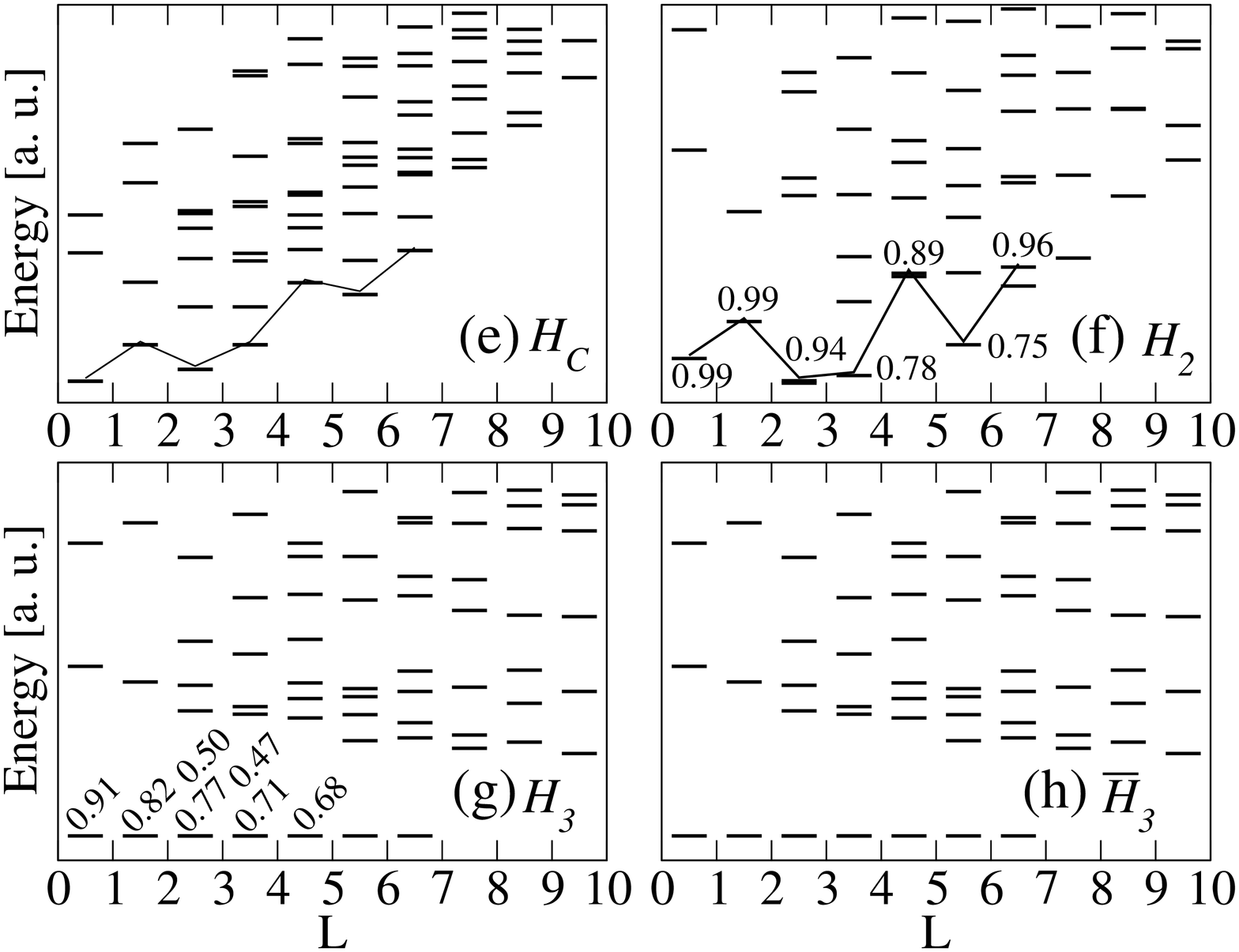}}\\
\mbox{\includegraphics[width=5.1cm,angle=-90]{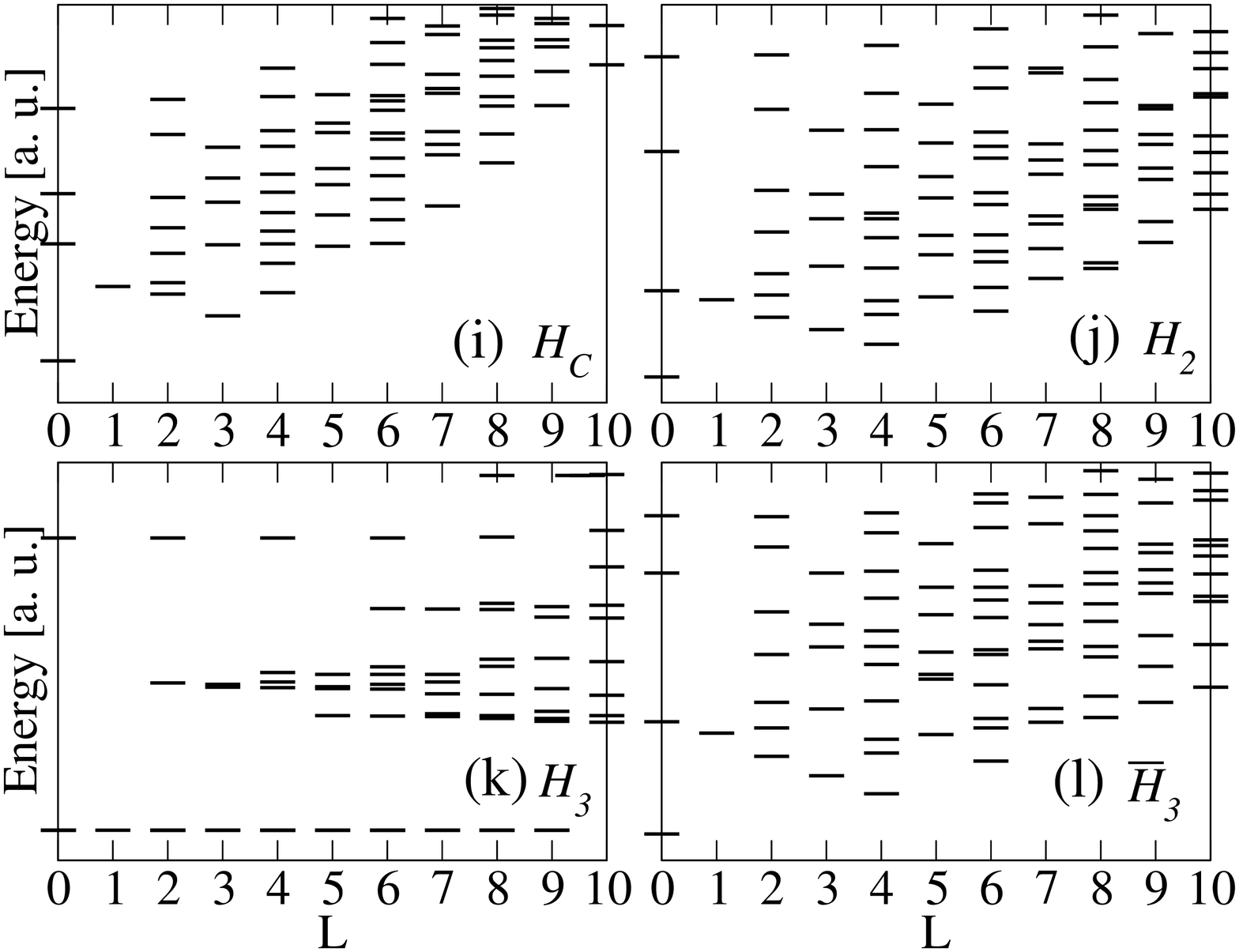}}
\end{center}
\caption{Exact energy (arbitrary units) as a function of total
angular momentum $L$ for a system with $N_e=8$ [(a)-(d)],
$N_e=7$ [(e)-(h)], and $N_e=6$ [(i)-(l)] at $N_\phi=13$ for the
four considered Hamiltonians, $H_C$, $H_2$, $H_3$, and
$\overline{H}_3$. Numbers inside the plots are the wave function
overlaps between the indicated eigenstates and the lowest energy
eigenstates (at that $L$) of $H_C$.  The lines connecting the
low energy states in (e) and (f) are a guide to the eye. In
(g) two numbers are given at $L=1.5$ and $2.5$ for the 
two degenerate ground states at those angular momenta.}
\label{spectra}
\end{figure}

Meanwhile, the relationship on the sphere between $N_\phi$ 
and $N_e$ for the Pf is
$N_\phi=2N_e-3$ where $-3$ is the ``shift''.  Since
$N_e$ is related to the number of holes $N_h$ through $N_h + N_e
= N_\phi+1$ the relationship between $N_e$ and $N_\phi$
for the \apf is $N_\phi=2N_h-3=2N_e+1$ and thus the Pf  
and \apf have different ``shifts''. On the torus,
PH symmetrizing the Pf creates a new state with a significantly
improved overlap with the exact Coulomb ground state in the
SLL~\cite{hr}. By contrast, on the sphere, such an
attempt obviously fails since PH symmetrizing the Pf
changes the particle number for a given flux. However, this
dichotomy between the $(N_\phi, N_e)$ relationships for the Pf and
\apf provides a convenient platform for addressing the issue of
spontaneous PH symmetry breaking. The reason is as follows.

The Pf and \apf break PH symmetry differently.  
In principle, the difference can be
parameterized in terms of an Ising-like ``order parameter'' measuring
the ``deviation'' from a PH-symmetric ground state. Fortunately,
on the sphere, the Pf and \apf belong to different
$N_e$ sectors for a given flux and $N_e$ can be
regarded an order parameter; $N_e=(N_\phi+3)/2$ for the Pf and
$N_e=(N_\phi-1)/2$ for the \apf with a
PH-symmetric ground state obtained exactly in the middle.  
We conclude that the ground state energy can be
regarded as the ``Landau free energy'' with $N_e$ being an order
parameter (at least so long as $N_e$ does not deviate too far
from half filling). The question then becomes whether this ``Landau
free energy'' exhibits a Mexican-hat structure for $H_2$ which,
if present, would indicate spontaneous PH symmetry breaking.
Another important and experimentally relevant question is what
happens in the case of the Coulomb interaction both with and
without finite-thickness effects.

\begin{figure*}[t]
\begin{center}
\mbox{\includegraphics[width=4.5cm,angle=-90]{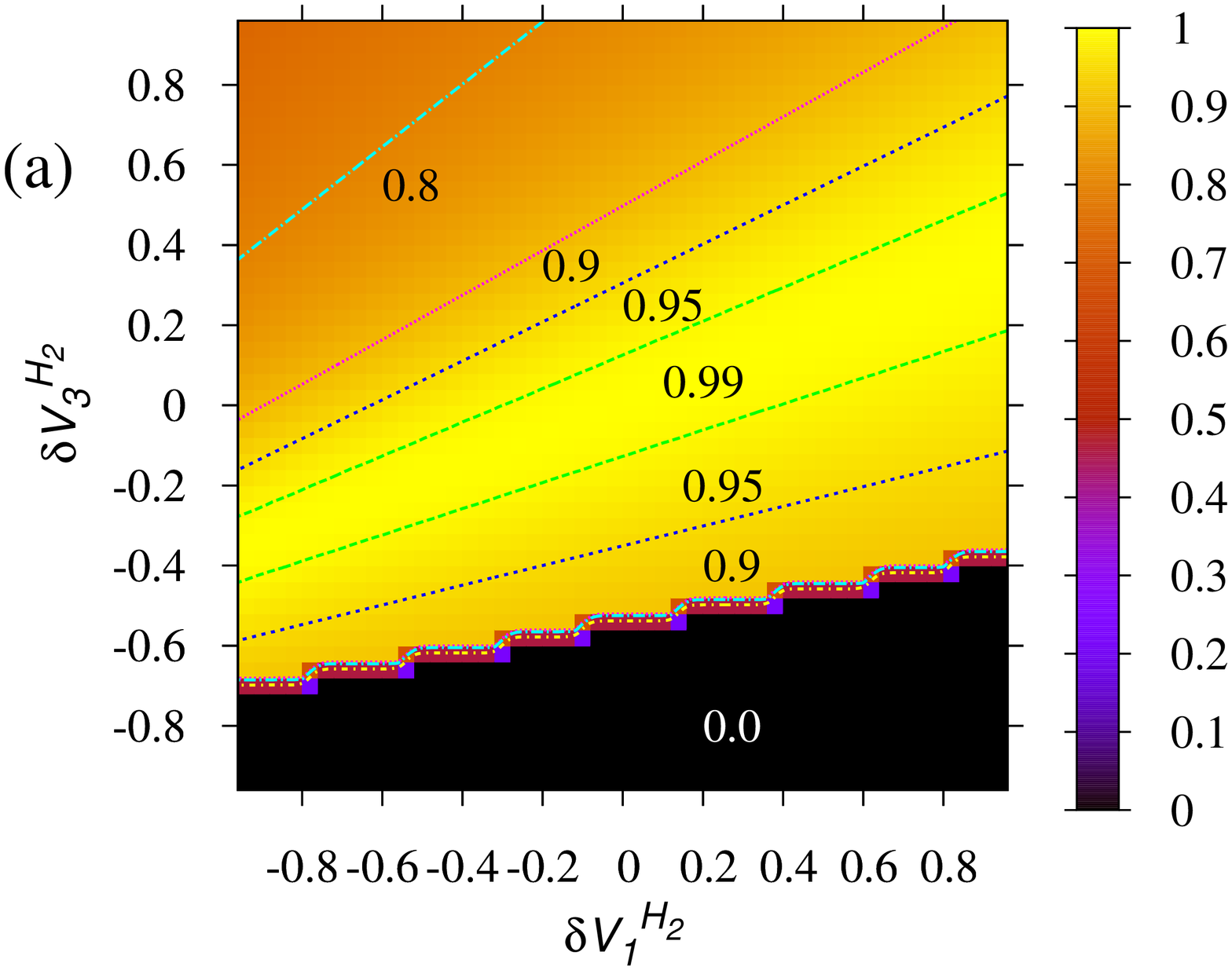}}
\mbox{\includegraphics[width=4.5cm,angle=-90]{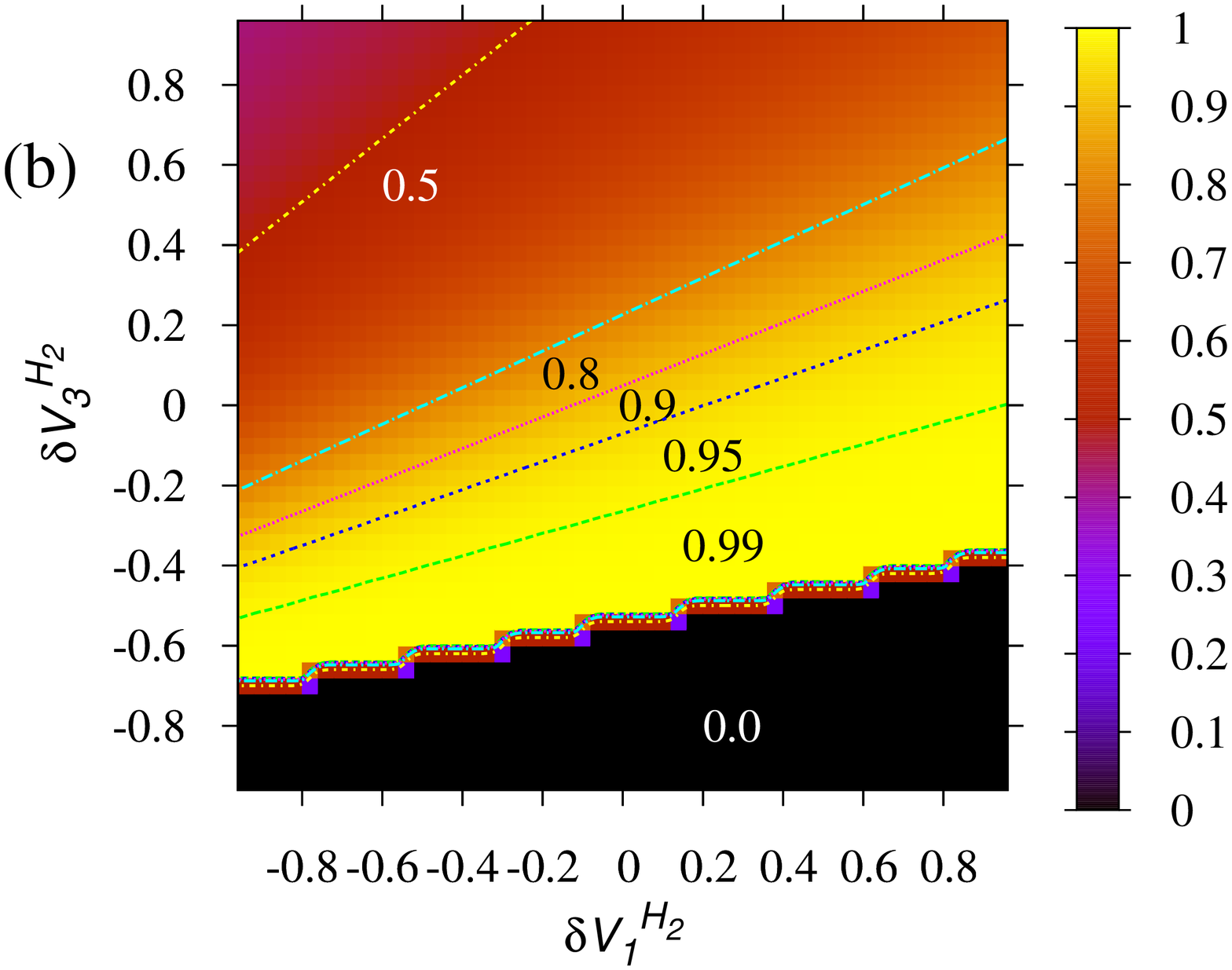}}
\mbox{\includegraphics[width=4.5cm,angle=-90]{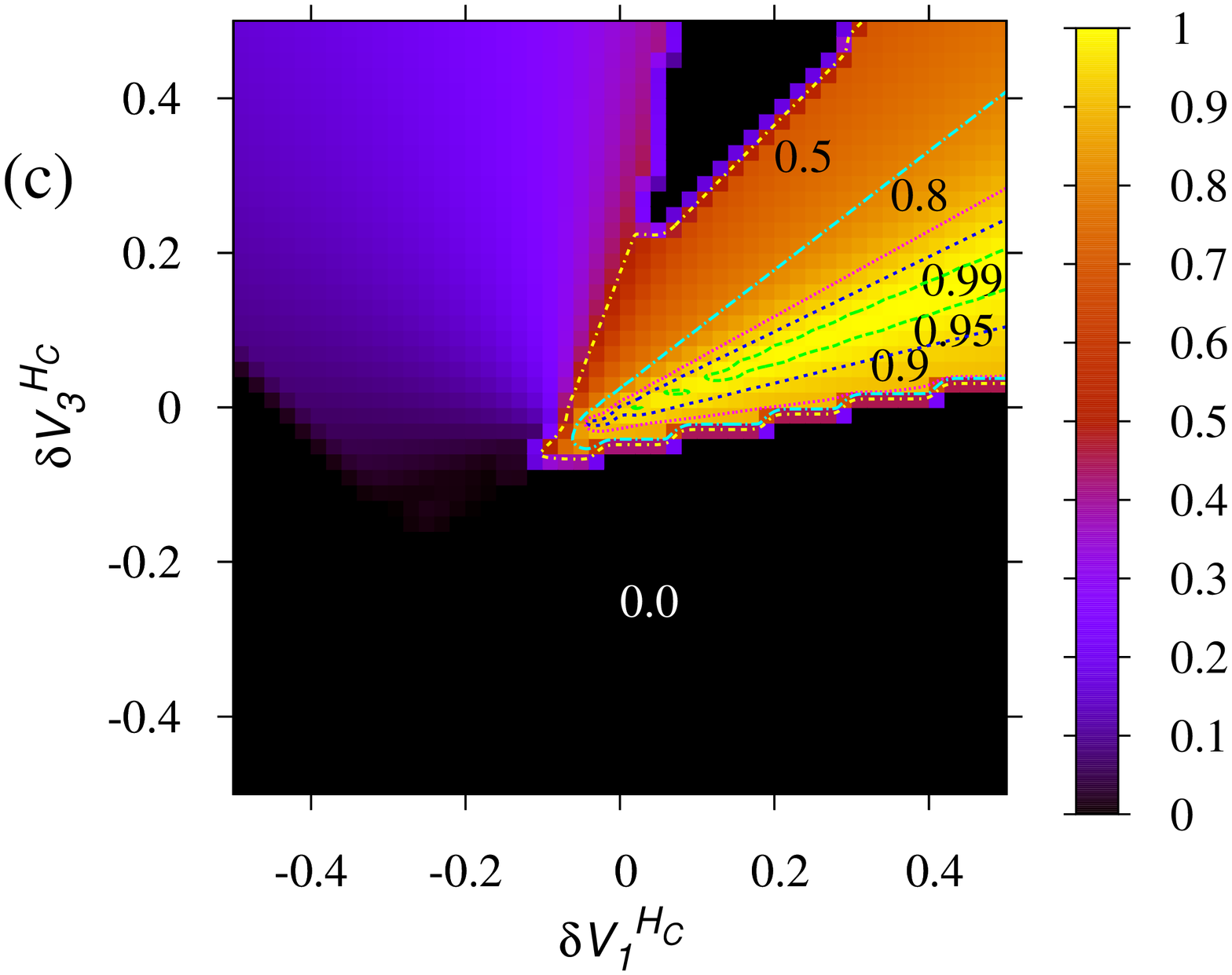}}
\mbox{\includegraphics[width=4.5cm,angle=-90]{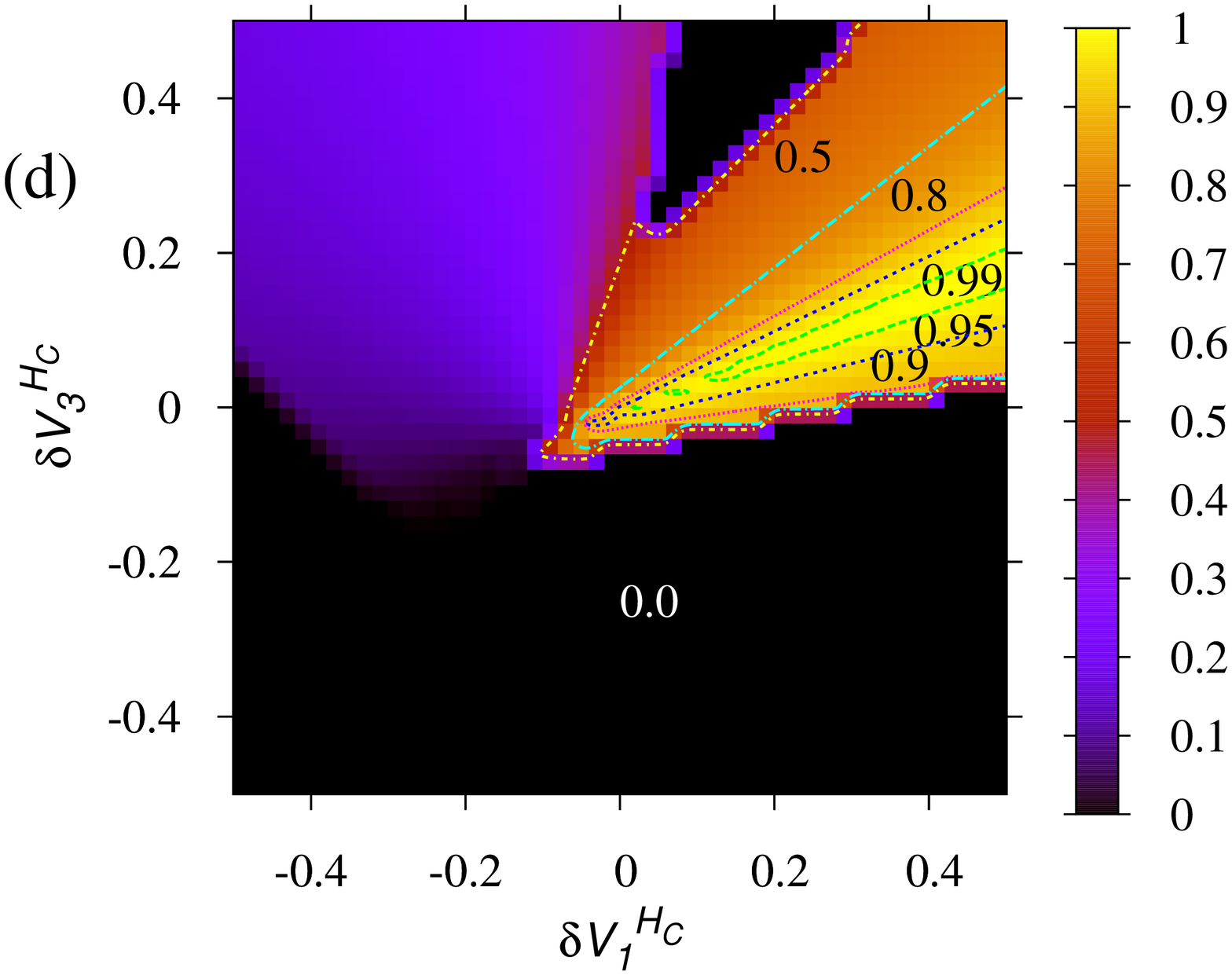}}
\end{center}
\caption{(Color online) Contour plot of (a) $\langle
\Psi^{H_3}_0|\Psi^{H_2}_0\rangle$ perturbing $\delta V^{H_2}_1$
and $\delta V^{H_2}_3$, (b) $\langle
\Psi^{H_C}_0|\Psi^{H_2}_0\rangle$ perturbing $\delta V^{H_2}_1$
and $\delta V^{H_2}_3$, (c) $\langle
\Psi^{H_C}_0|\Psi^{H_2}_0\rangle$ perturbing $\delta V^{H_C}_1$
and $\delta V^{H_C}_3$, and (d)  $\langle
\Psi^{H_C}_0|\Psi^{H_3}_0\rangle$ perturbing $\delta V^{H_C}_1$
and $\delta V^{H_C}_3$.} \label{contour}
\end{figure*}

Before addressing the above questions we
confirm that $H_2$ is indeed a good model
Hamiltonian containing the physics of $H_3$. To
this end, consider $H_2$ at the Pf value of $N_e$ for a given
$N_\phi$, $N_{Pf}= (N_\phi+3)/2$, and change the value of $N_e$
from $N_{Pf}$ to the \apf value, $N_{\overline{Pf}}=
(N_\phi-1)/2$. At $N_{Pf}$, $H_3$ produces energy spectra
similar to those of the SLL $H_C$ (denoted only $H_C$ from this point) 
and generates the Pf as the ground
state; see Fig.~\ref{spectra}(a) and (c), where $N_\phi=13$. In
contrast, $\overline{H}_3$ creates \apf quasiparticles
forming a degenerate ground state manifold shown in Fig.~\ref{spectra}(d). 
Important for our purpose is that the low-energy
spectra of $H_2$ (Fig.~\ref{spectra}(b)) is essentially equivalent to that of $H_3$ in
this number sector. At $N_{\overline{Pf}}$, the role of the Pf and
\apf are exactly reversed where $H_2$ generates almost the
identical low-energy spectra as $\overline{H}_3$ and 
also $H_C$; see Fig.~\ref{spectra}(i)-(l).

Results between $N_{Pf}$ and $N_{\overline{Pf}}$ are most
intriguing. Due to the exact PH symmetry, $H_3$ and
$\overline{H}_3$ have precisely the same energy spectra
containing a degenerate manifold of Pf quasiholes and \apf
quasiparticles; see Fig.~\ref{spectra}(g) and (h). A surprising
fact is that $H_2$ produces an energy spectra, shown in 
Fig.~\ref{spectra}(f), qualitatively
similar to those of $H_C$ as a result of an intricate
interaction between an equal mixture of 
Pf quasiholes and \apf quasiparticles. Quantitatively, the similarities can be
investigated via the wave function overlaps calculated for the lowest
branch of excitations ranging from $0.75$ to $0.99$. These are
compared to those of $H_3$ (or $\overline{H}_3$) which are usually
below 0.9 and predominantly lower.

Since $H_2$ is a two-body interaction Hamiltonian, it is useful to
calculate the Haldane pseudopotentials (on the sphere) through
$H_2 = \sum_m V^{N_\phi}_{m}\sum_{i<j} {\cal P}_{ij}
(N_\phi-m)$, where ${\cal P}_{ij}(L)$ projects onto
states with relative pair angular momentum $L$.  The
pseudopotentials on the sphere are connected to those
on the plane through the thermodynamic limit:
$V_m=\lim_{N_\phi\rightarrow\infty}V^{N_\phi}_{m}$.  
Explicit computation finds that, surprisingly,
$V^{N_\phi}_{m}$ is non-zero only for $m=1$ and $3$ and
the interaction is quite short-ranged and nearly hardcore
compared to the Coulomb interaction. For the sake of further
theoretical studies we provide the planar-geometry Haldane
pseudopotentials: $V^{H_2}_1=2.7119(10)$ and
$V^{H_2}_3=0.90173(73)$ where the numbers in parentheses represent
the statistical error from taking the thermodynamic limit. (The 
superscript $H_2$ denotes the Hamiltonian from which the 
pseudpotentials $V^{H_2}_m$ are derived.)

To test the robustness of the Pfaffian-like description of the
ground state of $H_2$ we consider $N_e=8$ electrons at
$N_\phi=13$ and diagonalize $H_2$ allowing $V^{H_2}_1$ and
$V^{H_2}_3$ to be perturbed from the original values.
Fig.~\ref{contour}(a) shows the overlap between the Pf state
$|\Psi_0^{H_3}\rangle$ and the exact $H_2$ ground state 
$|\Psi_0^{H_2}\rangle$ as a function of the variations in the
$H_2$ pseudopotentials $\delta V^{H_2}_1$ and $\delta V^{H_2}_3$.
$|\Psi_0^{H_2}\rangle$ remains Pf-like to a large
degree over a significant parameter range. A similar ``phase''
diagram can be obtained for the overlap between
$|\Psi_0^{H_2}\rangle$ and the $H_C$ ground state 
$|\Psi_0^{H_C}\rangle$ where the $H_2$ pseudopotentials are
allowed to vary, cf., Fig.~\ref{contour}(b). Again, 
there is a large region of the $\delta V^{H_2}_1$-$\delta
V^{H_2}_3$ plane where $|\Psi_0^{H_2}\rangle$ approximates
$|\Psi_0^{H_C}\rangle$ very accurately. Finally, we compute
the overlap between $|\Psi_0^{H_2}\rangle$ and
$|\Psi_0^{H_C}\rangle$  perturbing the first two
pseudopotentials of  $H_C$ shown in Fig.~\ref{contour}(c).
Here the overlap ``phase'' diagram is nearly identical to that
obtained between the Pf and the Coulomb ground state--the
difference between the two [Figs.~\ref{contour}(c) and (d)] is
always less than $2.5 \%$. Note that, for positive $\delta
V^{H_C}_1$ and $\delta V^{H_C}_3$, there is a region where the
ground state of  $H_C$ is well approximated by that of $H_2$
with the same being true for $H_3$~\cite{morf,ft-long,hr}.

\begin{figure}[t]
\begin{center}
\mbox{\includegraphics[width=6.5cm,angle=-90]{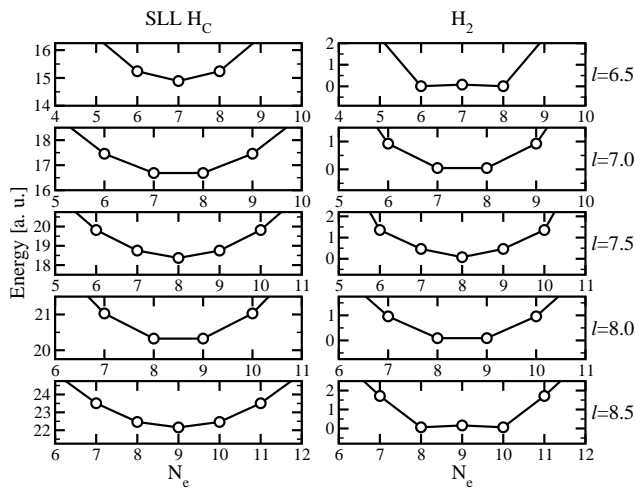}}
\end{center}
\caption{Comparison between the ground state energy of $H_C$
and $H_2$ as a function of electron number $N_e$ in the vicinity
of half filling. The total magnetic flux piercing the sphere is
$N_\phi=2(l-1)$ for $H_C$ and $2l$ for $H_2$. Note that an
appropriate chemical potential energy is added in the case of
$H_C$~\protect \cite{foot1} . }
\label{HC_vs_H2}
\end{figure}

We now present our main results.  Figure~\ref{HC_vs_H2} shows 
a comparison between the ground state energy of $H_C$ 
and $H_2$ as a function of $N_e$ in the
vicinity of half filling for various magnetic fluxes. The ground state energy 
of $H_C$ is essentially 
featureless except an overall parabolic envelope which is nothing
but the charging energy of a finite-size system. In contrast,
$H_2$ exhibits salient ``Mexican-hat'' structures whenever the Pf
and \apf occur at even particle numbers given by the respective
$(N_\phi,N_e)$ relationships: see $l=6.5$ and $8.5$ in
Fig.~\ref{HC_vs_H2}. It is important to note that when nominal
particle numbers for the Pf and \apf become odd at $l=7.5$ the
``Mexican-hat'' structure disappears.

Therefore, it is shown in our numerical studies that the PH
symmetry is likely to be spontaneously broken in $H_2$, but not 
in $H_C$.  We emphasize that the 
existence of the ``Mexican-hat''
structure in finite-size systems is usually a necessary condition
for spontaneous PH symmetry breaking in the thermodynamic limit.
Also, in our further numerical studies the
``Mexican-hat'' structure remains absent even when
finite-thickness effects are incorporated 
into  $H_C$ (or for the zero width LLL $H_C$). It is
interesting that, despite the high overlaps between the ground
states of $H_2$ and  $H_C$ at the Pf (\apf) sector, energy
landscapes of the two Hamiltonians become qualitatively different
in the vicinity of half filling.

Finally we mention experimental implications. Since
the PH symmetry is not spontaneously broken in the case of the
Coulomb interaction, with or without finite-thickness, 
it is likely that the true ground state is neither the
pure Pfaffian nor anti-Pfaffian state in the absence of external
PH-symmetry breaking terms such as those inherent with LL mixing. While it is possible
that large LL mixing favors one state over another, a possible
scenario is that the Pf and \apf are linearly superposed in the
thermodynamic limit (where the number difference between the two
states is infinitesimal compared to the total particle number) and
form a new PH-symmetric ground state, in which case the edge-state
behavior would be quite different from that of the pure 
Pf or \apf~\cite{apf1,apf2}.    Another possibility is that local disorder could induce spatially 
random LL mixing producing spatially random local PH-symmetry breaking.  This would yield 
random domains of Pf and \apf states in the two-dimensional plane.  Either scenario is consistent with
the previously mentioned fact that the overlap with the Coulomb
ground state is significantly improved in the torus geometry when
PH symmetrization is explicitly applied to the Pf.   The other possibility that the ``hat 
structure'' is actually present in the Coulomb case, with an effect too small 
to been seen in numerical work, is unlikely in our opinion.

MRP and SDS acknowledge support from Microsoft Project Q. KP would
like to thank Asia Pacific Center for Theoretical Physics (APCTP)
for its hospitality.




\begin{thebibliography}{99}
\bibitem{foot} The Moore-Read Pfaffian state decribes
fully spin polarized electrons consistent with the current
theoretical~\cite{morf,feiguin2} and experimental consensus 
and we assume full spin-polarization of the 
5/2 FQHE although recent experiments~\cite{dean} raise 
some interesting questions on this issue.
\bibitem{foot1}We consider $(H_C + \overline{H}_C)/2$ (where $\overline{H}_C$ is the 
PH conjugate to $H_C$) to ensure that the energy spectra (not just eigenstates) 
for $N_e$ electrons is equivalent to $N_\phi+1-N_e$ electrons.  This is equivalent
to adding an appropriate $N_\phi$ and $N_e$ dependent (for finite systems) chemical 
potential to $H_C$.
\end{thebibliography}
\end{document}